\documentstyle[aps,prl,epsf,psfig]{revtex}
\begin{document}


\draft
\preprint{\today}
\title{Influence of wetting properties on hydrodynamic boundary
conditions at a fluid-solid interface}

\author{Jean-Louis Barrat }
\address{ D\'epartement de Physique des Mat\'eriaux \\
Universit\'e Claude Bernard and CNRS, 69622 Villeurbanne Cedex, France}

\author{Lyd\'eric Bocquet}
\address{Laboratoire de Physique\\
ENS-Lyon and CNRS, 69364 Lyon Cedex 07, France}

\date{\today}

\maketitle

\begin{abstract}
It is well known that, at a macroscopic level, the boundary condition
for a viscous fluid at a solid wall is one of "no-slip".
The liquid velocity field vanishes at a fixed solid boundary. 
In this paper, we consider the special case of a liquid that partially
wets the solid, i.e. a drop of liquid in equilibrium
with its vapor on the solid substrate has a finite contact angle.
Using extensive Non-Equilibrium Molecular Dynamics (NEMD) simulations, we show that when the contact angle is large enough, the boundary condition can drastically differ (at a microscopic level) from a "no-slip" condition.
Slipping lengths exceeding 30 molecular diameters are obtained 
for a contact angle of 140 degrees, characteristic of Mercury on glass. 
On the basis of a Kubo expression for $\delta$, we derive an expression for 
the slipping length in terms of equilibrium quantities of the system.
The predicted behaviour is in very good agreement with the  
numerical results for the slipping length obtained in the NEMD simulations.
The existence of large slipping lentgh may have important implications for the transport properties
in nanoporous media under such "nonwetting" conditions.

\end{abstract}

\pacs{Pacs numbers: 61.20J 68.15 68.45G}

\section{Introduction}

Properties of confined liquids have been the object of constant
interest during the last two decades, thanks to the considerable 
development of Surface
Force Apparatus (SFA) techniques. While static properties are
now rather well understood (see e.g. \cite{IS85}), the 
dynamics of confined systems have been investigated more recently \cite{Klafter}. 
These studies have motivated
much numerical and theoretical work 
\cite{KB88,TR90,BS90,BB94,Mundy,Koplik} and
some progress has been made in giving a simple coherent description of the
collective dynamics of confined liquids. Both from
experimental and theoretical studies has emerged a rather simple
description of the dynamics of not too thin liquid films -i.e. 
films thicker than typically 10 to 20 atomic sizes. 
The hydrodynamics of the film can be described 
by  the macroscopic hydrodynamic equations with bulk transport coefficients, 
supplemented by a "no-slip" boundary condition applied in the 
vicinity (i.e. within one molecular
layer) of the solid wall. Hence, in spite of the fact that the wall induces
a structuration of the fluid into layers that can extend 5 to 6 
molecular diameters from the wall, the hydrodynamic properties of the 
interface are quite simple.

It turns out, however, that all experimental and numerical studies
of confined fluids that  have been 
carried out with fluid/substrate combinations that correspond to 
a total wetting  situation. By this we mean that $\gamma_{LS} + \gamma_{LV} < \gamma_{SV}$, where $\gamma$ is a surface energy and the indexes
$L$,$V$ and $S$ refere to the liquid, its vapor and the substrate, 
respectively \cite{note1}. In this letter, we investigate the structure 
and the  hydrodynamic  properties of a fluid film that
is forced to penetrate a narrow pore in a  situation of partial wetting, i.e.
when $\gamma_{LS} + \gamma_{LV} > \gamma_{SV}$.  This corresponds to the case
where a drop of the liquid resting on the same substrate, at equilibrium with
its vapor, has a finite contact angle which can be deduced from Young's law
$\gamma_{LV}\cos \theta= (\gamma_{SV}-\gamma_{LS})$ \cite{RW}.

\section{Model and Results}

We first describe our model for the fluid and the substrate, and 
some details of the simulation procedure. All interactions 
are of the Lennard-Jones type, 
\begin{equation}
v_{ij}(r) = 4 \epsilon \left( \left({\sigma\over r}\right)^{12} - c_{ij}
\left({\sigma\over r}\right)^6 \right)
\label{e1}
\end{equation}
with identical interaction energies and molecular diameters $\sigma$.
The surface energies will be adjusted by tuning the coefficients
$c_{ij}$. In all the simulations that are presented in this letter,
the solid substrate is described by atoms fixed on a FCC lattice, 
with a reduced density $\rho_S \sigma^3 = 0.9$. As the atoms are
fixed, the coefficient $c_{SS}$ is in fact irrelevant. The interactions 
between fluid atoms are characterized by $c_{FF}=1.2$, meaning that the
fluid under study is more cohesive than the usual Lennard-Jones fluid. 
The fluid-substrate interaction coefficient $c_{FS}$ will be varied between 
$0.5$ and $1$. All the simulations will be carried out at a constant reduced 
temperature, $k_B T/\epsilon = 1$. 

Finally, we mention
that the   configuration under study will be that of an 
 fluid slab confined between two parallel solid walls.
Typically, a configuration contains $10000$ atoms, with a distance between 
solid walls $h=18\sigma$ and lateral cell dimensions 
$L_x=L_y=20\sigma$. Periodic boundary conditions are applied in the $x$
and $y$ directions, i.e. 
parallel to the walls. For each wall, three layers  of FCC
 solid  (in the 100 orientation) will be
modelled using point atoms, a continuous attraction between the 
fluid and the wall in the direction perpendicular to the walls 
being added  in order to model the influence of the deeper solid layers.

All simulations were carried out at constant temperature ($k_BT/\epsilon =1$)
by coupling the fluid atoms to a Hoover's thermostat \cite{AT87}. In flow
 experiments,
only the velocity  component  in the direction orthogonal to the flow was 
thermostatted.

Before we discuss in detail the structure of a film, we can
roughly estimate the influence of the interaction parameters
on the wetting properties of the fluid. Following the standard
Laplace estimate of surface energies \cite{RW}, we have
$\gamma_{ij} = -\rho_i\rho_j \int_{r_0}^\infty r u_{ij}(r) dr$.
Using Young's law, we obtain for the contact angle
$\cos\theta = -1 + 2\rho_S c_{FS} /\rho_F c_{FF}$. 

\begin{figure}
\psfig{file=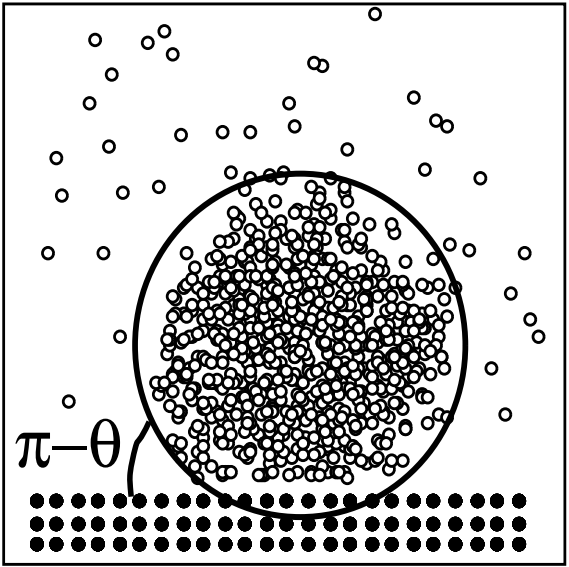,width=7.5cm,height=7.5cm}
\vspace{6mm}
\caption{A typical configuration of a liquid droplet (1000 atoms)
on a solid substrate in a "nonwetting" case ($c_{FS}=0.5$),
in equilibrium with its vapor.
This configuration was prepared by starting from an homogeneous
fluid slab  of thickness $18\sigma$, confined between two walls.
The droplet is formed by simultaneously increasing moving the upper wall 
by $20\sigma$ in the $z$ direction and removing the fluid atoms that
lie near the box boundaries. The bounding box has a size of $20\sigma$.
}
\label{fig1}
\end{figure}

From this expression a variation of $c_{FS}$ between 0.5 and 0.9
would be expected to induce a variation of $\theta$ between 100 degrees and
50 degrees.  
A more accurate  determination of the surface tensions
was  carried out  using the method of Nijmeijer {\it et al.} \cite{Nijmeijer}
The surface tensions are defined in terms of an integral over components of the pressure tensors which can be computed in a simulation. We refer to
ref. \cite{Nijmeijer} for more details. The results 
are listed in table I. By tuning the solid-fluid interaction strength $c_{FS}$ from $c_{FS}=1.0$ to $c_{FS}=0.5$, we found 
the contact angle deduced from Young's law varies from 
$\theta\simeq 90$ degrees to $\theta\simeq140$ degrees. 
In figure \ref{fig1}, a typical configuration
of a liquid droplet (in coexistence with its vapor)
on the solid substrate is shown for $c_{FS}=0.5$, corresponding to
$\theta\simeq140$ degrees. In the following we shall describe 
such large contact angles (i.e. larger than 90 degrees)
as corresponding to  a "nonwetting" situation.

In order to force the  fluid into a 
narrow liquid pore under such partial wetting conditions,
 an external pressure has to be applied. A simple
thermodynamic argument shows that for a parallel slit
of width $h$, the minimal pressure is $P_0 = 2(\gamma_{LS} -\gamma_{SV})/h$.
For the fluid with $c_{FS}=0.5$, we find, for $h=18\sigma$,  $P_0=0.079 \epsilon/\sigma^3 $, while $P_0= 0.018 \epsilon/\sigma^3$ when
$c_{FS}=0.9$. If we use $\sigma=5$\AA, $\epsilon=0.05$eV, then $P_0 \sim$MPa for 
$h=9$nm in the "nonwetting" case, $\theta = 140$ degrees.

\begin{figure}
\psfig{file=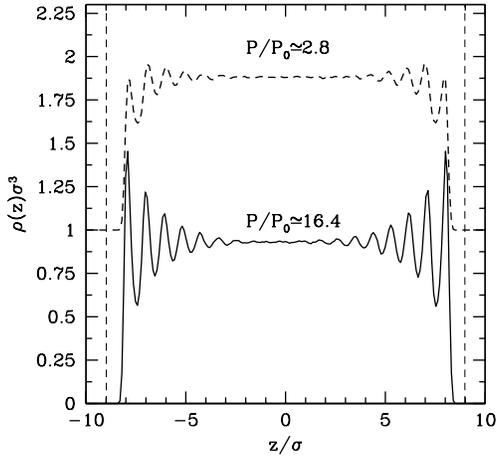,width=7.5cm,height=7.5cm}
\vspace{6mm}
\caption{Density profiles of the "nonwetting" fluid
($c_{FS}=0.5$) confined between two solid walls 
separated by 20$\sigma$. The positions of the first layer
of solid atoms have been indicated by vertical dashlines.
Full line: $P/P_0=2.8$; dashed line: $P/P_0=16.4$. The latter curve
has been shifted upwards for clarity.}
\label{fig2}
\end{figure}

 Figure \ref{fig2} shows the density profiles of the
nonwetting fluid inside the pore for pressures corresponding to
$2.8 P_0$ and $16.4 P_0$. The pressure is changed at
 constant pore width by changing
the number of particles. It is seen in this figures that the highest pressure
structure strongly resembles what would
be obtained for the usual case of a wetting fluid, with 
a strong layering at the wall. The structure at the lower pressure is markedly different, with both a layering parallel to the wall and
a density depletion near the wall.

\begin{figure}
\psfig{file=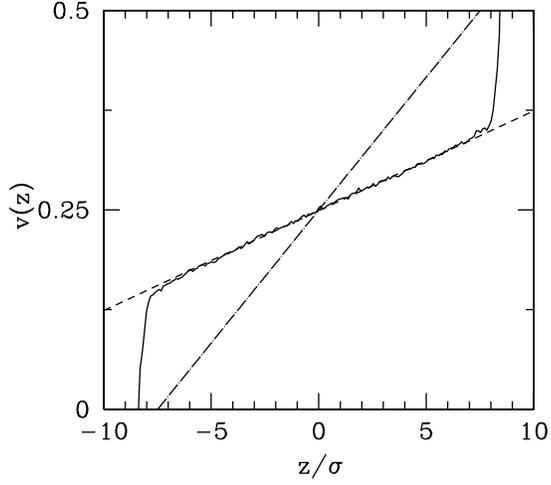,width=7.5cm,height=7.5cm}
\vspace{6mm}
\caption{Velocity profile (in reduced units) of the "nonwetting" fluid ($c_{FS}=0.5$) in 
a Couette geometry. The reduced pressure is $P/P_0\simeq 7.3$. The solid line is the simulation result, the dashed
line is a linear fit of the numerical results, and the dashed-dot line is the velocity profile
predicted by the no-slip bounbary condition. The velocity of the upper wall is
$U=0.5$.  }
\label{fig3}
\end{figure}

\begin{figure}
\psfig{file=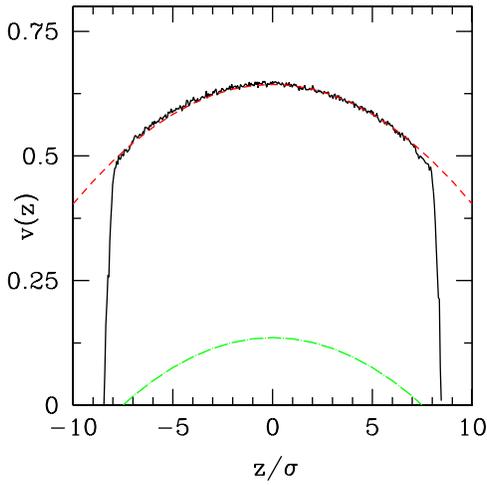,width=7.5cm,height=7.5cm}
\vspace{6mm}
\caption{Same as in figure 3, but in a Poiseuille geometry. The external
force applied on each fluid particle is $f_{ext}=0.02 \epsilon/\sigma$ 
}
\label{fig4}
\end{figure}

We now turn to the study of the dynamical properties 
of the confined fluid layer. Two types of numerical experiments,
corresponding to Couette and Poiseuille planar flows were
carried out. In the Couette flow experiments, the upper wall 
is moved with a velocity $U$ (typically $U=0.5$ in reduced Lennard-Jones units).
In the Poiseuille flow experiment, an external force in the $x$ direction is applied to the fluid particles. In figures \ref{fig3} and \ref{fig4} we compare the resulting
velocity profiles  to  those that would be expected for a "no slip" 
boundary condition applied at one molecular layer from the solid wall.
Obviously the velocity profiles for the nonwetting
fluid imply a large amount of slip at the solid boundary. As usual,
this slippage effect can be quantified by introducing
a "partial slip" boundary condition for the tangential velocity $v_t$
at the solid liquid boundary:
\begin{equation}
{\partial v_t  \over \partial z}|_{z=z_{w}} = -
 {1 \over \delta} v_t|_{z=z_{w}}  .
\label{def_slip}
\end{equation}
This boundary condition depends on two parameters, the wall
location $z_w$ and the slipping length $\delta$. By studying
simultaneously Couette and Poiseuille flow for the same
fluid film, both parameters can be determined
if they are used as fit parameters for the velocity profiles
obtained in the simulation.  The results of such an adjustment
are shown in figures \ref{fig3} and \ref{fig4}.
It turns out that, as was the case in earlier studies
\cite{BB94}, the hydrodynamic position $z_w$ of the walls 
is located inside the fluid, typically one atomic distance
from the outer layer of solid atoms.  Much more interesting is the
variation of the slipping length $\delta$, which in earlier work
was always found to be very small.  In figure \ref{fig5},
the variation of $\delta$ as a function of the pressure
is shown for several values of the interaction parameters.
The pressure are normalized by the capillary pressure 
$P_0$ defined above as the minimal pressure that must be
applied to the fluid in order to enter the pore. For an interaction
parameter $c_{FS}=0.9$, corresponding to a contact angle
$\theta = 100$degrees , the usual behaviour (i.e. a small $\delta$) is obtained. For an
interaction parameter $c_{FS}=0.5$ which corresponds to a contact
angle $\theta=150$ degrees \cite{note2}, slipping lengths larger than $15$ molecular diameters
can be obtained at the lowest pressures; even at relatively 
high pressures ($10 P_0$), the slipping length remains appreciably larger
than the molecular size $\sigma$.

\section{Theory}
In order to understand the relation between the hydrodynamic boundary condition
and the wetting properties of the fluid on the substrate, one needs to estimate
the dependence of the slipping length $\delta$ on the microscopic parameters
of the system, such as ``roughness'', temperature, $c_{FS}$, etc...

From the ``kinetic'' point of view, this is obviously a hard task to complete 
since the slipping length accounts for the `parallel' transfer of momentum between 
the fluid and the substrate. Explicit calculations can only be done in some
model systems \cite{CRAS}. Now in the general case of a dense fluid,
no explicit formula for $\delta$ in terms of  microscopic quantities is available in the litterature. 

In the following, we derive an approximate expression for the slipping length,
which will allow us to discuss qualitatively the relationship 
between wetting and boundary conditions.

Our starting point is a Green-Kubo expression for the slipping length \cite{BB94} :
\begin{equation}
\lambda={\eta \over \delta} = {1\over {Ak_BT}} \int_0^{+\infty} \langle F_x(t).F_x(0) 
\rangle ~dt
\label{GK}
\end{equation}
where $\eta$ is the shear viscosity of the fluid, $A=L_xL_y$ is the lateral solid
surface, and $F_x$ is the x-component of the total force due to the wall 
acting on the fluid, at equilibrium (x is a component parallel to the wall).
The quantity
$\lambda=\eta/\delta$ can be interpreted as the friction coefficient of
the fluid/wall interface, relating the force along x due to the wall, to the
fluid velocity slip at the wall 
\begin{equation}
\langle F_x \rangle = -A~\lambda~V_{slip}
\label{lambda}
\end{equation}
  
By introducing the density-density correlation function, eq. (\ref{GK})
can be rewritten
\begin{equation}
{\eta \over \delta} = {1\over {Ak_BT}} \int_0^{+\infty} dt
\int d{\bf r}_1\int d{\bf r}_2~F_x({\bf r}_1)~F_x({\bf r}_2)~
\langle\rho({\bf r}_1;t)~\rho({\bf r}_2;0))\rangle
\label{eq1}
\end{equation}
The force field $F_x({\bf r}_1)$ derives from the fluid-substrate potential
energy $V(xyz)$. The latter has been computed by Steele for a periodic
substrate interacting with the fluid through Lennard-Jones interactions
\cite{Steele}. In the case of a (100) face, the main contribution can be 
written in terms of the shortest reciprocal lattice vectors according to
\begin{equation}
V(xyz)=V_0(z)+ V_1(z) \left\{ \cos \left(q_\parallel x\right) +
\cos \left(q_\parallel y\right) \right \}
\label{pot}
\end{equation}
where $q_\parallel=2\pi/a_s$ and $a_s$ is the lattice spacing in the fcc solid.
The altitude $z$ is the distance to the first layer of the atoms of the
solid. The functions $V_0(z)$ and $V_1(z)$ are given by Steele in ref 
(\cite{Steele}). In our case however, the {\it attractive} parts of $V_0(z)$ and $V_1(z)$ are multiplied by the tuning factor $c_{FS}$.
The force along $x$ is the derivative with respect to
$x$ of the interaction potential $V(xyz)$ :
\begin{equation}
F_x(x,y,z) =q_\parallel~V_1(z)~\sin\left(q_\parallel x\right)
\label{fx}
\end{equation}
When inserted into eq. (\ref{eq1}), one obtains
\begin{equation}
{\eta \over \delta} = {1\over {Ak_BT}} \int_0^{+\infty} dt \int dx_1dy_1dz_1
\int dx_2dy_2dz_2~
q_\parallel^2 V_1(z_1) V_1(z_2)
\sin\left(q_\parallel x_1\right)\sin\left(q_\parallel x_2\right)
\langle\rho({\bf r}_1;t)~\rho({\bf r}_2;0))\rangle
\label{eq2}
\end{equation}
We now introduce the Fourier transform of the density in the plane
parallel to the substrate
\begin{equation}
\rho_{\bf k}(z)(t)=\int d{\bf x}~\rho({\bf x},z;t)~e^{i~{\bf k}\cdot{\bf x}}
\label{FT}
\end{equation}
with ${\bf x}=(x,y)$ and the vector ${\bf k}$ is parallel to the substrate.
This allows to rewrite the previous equation (\ref{eq2}) as
\begin{equation}
{\eta \over \delta} = {q_\parallel^2\over {2 A k_BT}}~ \int dz_1 \int dz_2~V_1(z_1)~V_1(z_2)~\int_0^{+\infty} dt~\Re \left(\langle \rho_{q_{\parallel}}(z_1)(t)
\rho_{-q_{\parallel}}(z_2)(0)\rangle \right)
\label{eq3}
\end{equation}
with $q_{\parallel}$ in the $x$ direction and $\Re$ stands for 
the real part.
Note that in deriving eq. (\ref{eq3}), the homogeneity of the system in the direction parallel to the substrate was taken into account.

Due to the presence of the confining solid, we now assume that the main 
contribution of $C(q_\parallel,z_1,z_2;t)$ to the time-integral comes from the dynamics in the plane $(x,y)$ parallel to the substrate. 
Moreover, these dynamics are probed at the first reciprocal lattice vector 
$q_\parallel$. Since $q_\parallel$ is close to the position of the first 
peak in the structure factor, it is reasonable to assume a diffusive relaxation
of $\langle \rho_{q_{\parallel}}(z_1)(t)
\rho_{-q_{\parallel}}(z_2)(0)\rangle$ \cite{deGennes,Boon}, yielding
\begin{equation}
\langle \rho_{q_{\parallel}}(z_1)(t)
\rho_{-q_{\parallel}}(z_2)(0)\rangle= exp(-q_\parallel^2~D_{q_\parallel} t) ~\langle \rho_{q_{\parallel}}(z_1) \rho_{-q_{\parallel}}(z_2)\rangle
\label{relax}
\end{equation}
where $D_{q_\parallel}$ is a {\it collective} diffusion coefficient
and $\langle \rho_{q_{\parallel}}(z_1) \rho_{-q_{\parallel}}(z_2)\rangle$ 
is the {\it static} correlation function.

The time integration can now be performed to obtain
\begin{equation}
{\eta \over \delta} \simeq {1\over {2 D_{q_\parallel}A k_BT}}~ \int dz_1 \int dz_2~V_1(z_1)~V_1(z_2)~\langle \rho_{q_{\parallel}}(z_1)
\rho_{-q_{\parallel}}(z_2)\rangle
\label{eq5}
\end{equation}
so that the slipping length is expressed in terms of static properties 
of the inhomogeneous system only. A further simplifications can be done
by assuming that, due to the stratification near the substrate, the main 
contribution in the integrals in eq. (\ref{eq5}) arises from the 
$z_1\simeq z_2$ terms, so that $ \langle \rho_{q_{\parallel}}(z_1)
\rho_{-q_{\parallel}}(z_2)\rangle \simeq
A\langle \rho (z_1) \rangle~ \delta (z_1-z_2) ~ S(q_{\parallel} | z_1)$.

The quantity $S(q_{\parallel} | z_1)$ is the $z$ dependent structure factor,
in the plane $z=z_1$ (parallel to the solid), defined as
\begin{equation}
S(q_{\parallel} | z_1)= {1\over {A\langle\rho (z_1)\rangle}}\langle \sum_{i,j}
e^{iq_{\parallel}(x_i-x_j)}~\delta(z_i-z_1)\rangle
\label{Sq}
\end{equation}
The factor $A\langle\rho (z_1)\rangle$ ($A$ being the lateral surface) normalizes
the average by the number of fluid particles in the layer at the
altitude $z_1$.
If no locking of the fluid occurs near the substrate, one may approximate $S(q_{\parallel} | z_1)$
by its value at the first layer, $S(q_{\parallel} | z_1) \simeq S_1(q_{\parallel})$. Equation (\ref{eq5}) thus reduces to
\begin{equation}
{\eta \over \delta} \simeq {S_1(q_{\parallel})\over {2 D_{q_\parallel}~k_BT}}~ \int dz_1~\rho (z_1)~V_1(z_1)^2
\label{eq6}
\end{equation}

In order to have a practical estimate to compare with, this formula can be further approximated. The integral term in eq. (\ref{eq6})
may be approximated by assuming that it is
dominated by the behaviour around the first layer located at $z_c\sim \sigma$,
so that 
\begin{equation}
\int dz~\rho (z)~V_1(z)^2 \sim \rho_c \int_\sigma^{+\infty} dz~ V_1(z)^2
\label{eq8}
\end{equation}
with $\rho_c$ the density at the first layer, denoted in the latter as
the ``contact'' density.
Moreover, one expects the ``long range'' attractive part of $V_1(z)$ to contribute mainly to the 
integral. Since the latter can be written $c_{FS}~V^{att}_1(z)$, with $V^{att}_1(z)$ independent of $c_{FS}$, one gets 
\begin{equation}
{\delta \over \sigma} \propto {D^*_{q_\parallel} \over {S_1(q_{\parallel}) ~ c_{FS}^2~\rho_c\sigma^3}}
\label{eq9}
\end{equation}
where $D^*_{q_\parallel}=D_{q_\parallel}/D_0$, and $D_0=k_BT/3\pi\eta\sigma$
is the Stokes-Einstein estimate for the bulk self diffusion coefficient.

All the quantities involved in eq. (\ref{eq9}) can be computed in 
equilibrium Molecular Dynamics simulations. The density at contact $\rho_c$
can be measured from density profiles, such as in fig. \ref{fig2}. On the
other hand, $D_{q_\parallel}$ and $S_1(q_{\parallel})$ can be computed 
from the correlations of density fluctuations in the first layer. In practice,
we introduce the function $S_1(q_\parallel,t)=N_1^{-1} \langle 
\rho_{q_\parallel}(t) \rho_{q_\parallel}(0) \rangle$, where $N_1$ is the
average number of particles in the first layer and $\rho_{q_\parallel}(t)=
\sum_{k=1,N_1} \exp iq_\parallel x_k(t)$ is restricted to atoms in the
first liquid layer. The value of $S_1(q_{\parallel},t)$ at time $t=0$
yields $S_1(q_{\parallel})$, while $D_{q_\parallel}$ is obtained in terms of the inverse relaxation time of $S_1(q_{\parallel},t)$, according to
eq. (\ref{relax}) . Let us note at this point that the assumption of
an exponential decay of $S_1(q_\parallel,t)$ is indeed verified in the
simulations, which allows us to clearly define $D_{q_\parallel}$.

\begin{figure}
\psfig{file=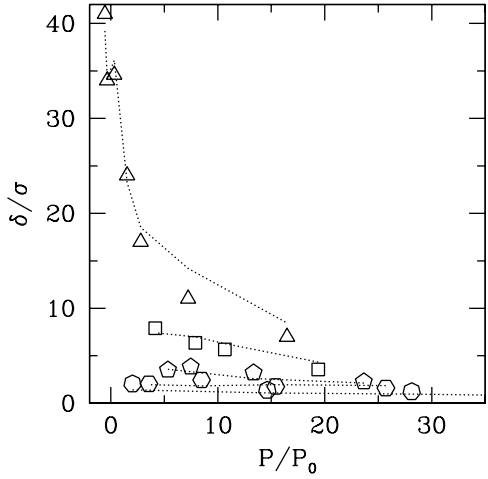,width=7.5cm,height=7.5cm}
\vspace{6mm}
\caption{Variation of the slipping length $\delta$ (in units of $\sigma$)
as a function of the reduced pressure $P/P_0$, for several values of the 
interaction parameters $c_{FS}$. From top to bottom, the datas correspond
to $c_{FS}=0.5$, $c_{FS}=0.6$, $c_{FS}=0.7$, $c_{FS}=0.8$, $c_{FS}=0.9$.
Solid lines are the theoretical prediction, eq. (\ref{eq9full}) (see text
for details).}
\label{fig5}
\end{figure}

In fig. \ref{fig5}, the ratio $\delta/\delta^*$, with $\delta^*/\sigma =
D^*_{q_\parallel} / {S_1(q_{\parallel}) c_{FS}^2}$ is plotted
as a function of the inverse density at contact, $1/\rho_c\sigma^3$.
In these variables, the theoretical estimate, eq. (\ref{eq9}), predicts
a linear dependence of $\delta/\delta^*$ as a function of $1/\rho_c\sigma^3$.
As shown in fig. \ref{fig5}, a linear behaviour $\delta/\delta^*=
\alpha (1/\rho_c\sigma^3 - 1/\rho_{shift}\sigma^3)$ is indeed observed,
in agreement with the prediction. A least-square fit of the datas in this plot
gives $\alpha=3.04$ and $1/\rho_{shift}\sigma^3=0.47$. The presence
of a shift in the density, $1/\rho_{shift}\sigma^3$, can be interpreted
to account for the higher-order correction in the density at contact
which have been neglected in deriving eq. (\ref{eq9}) (in particular
in the rough approximation assumed in eq. (\ref{eq8})). In the interesting limit where the contact density $\rho_c$ is small and $\delta$ is large, 
this shift does not contribute anymore. 

In fig. \ref{fig5}, the full theoretical result for $\delta$
\begin{equation}
{\delta \over \sigma} =\alpha {D^*_{q_\parallel} \over {S_1(q_{\parallel}) ~ c_{FS}^2~\rho_c\sigma^3}}\left(1-{\rho_c\over\rho_{shift}}\right)
\label{eq9full}
\end{equation}
is plotted as function of $P/P_0$ against the measured (out-of-equilibrium)
results for $\delta$. The good agreement obtained in these variables for
all different pressures and interaction strength $c_{FS}$ emphasize the
robustness of the previous theoretical estimate. Obviously, this expression
breaks down for very large contact density $rho_c>rho_{shift}=2.1 \sigma^3$,
where $\delta$ is expected to vanish anyway.

This result calls for several comments. First, eq. (\ref{eq9full}) shows that,
at for given fluid-substrate interaction $c_{FS}$, $\delta$
decreases with the density and structuration of the fluid in the first
layer. The slipping length is thus expected to be quite small in a dense
fluid at high pressures, as usually observed and measured experimentally
\cite{BB94,Georges}. More specifically, eq. (\ref{eq9full}) predicts
a strong dependence of $\delta$ on the value of the structure factor
in the first layer, taken at the shortest reciprocal lattice vector 
$q_\parallel$. This result is in qualitative agreement with previous
simulation results \cite{TR90}. 
Now if the fluid-substrate interaction $c_{FS}$ is decreased at a given 
contact density of the fluid $\rho_c$ (eg, by increasing simultaneously the pressure), the previous result predicts a strong increase of the slipping length.
This explains why substantial slip may be obtained, even if a strong
structuration does exist in the fluid, a fact which is {\it a priori} 
counter-intuitive.

Finally, let us come back to the problem of the influence of the wetting properties on the slipping length $\delta$. As noted above, eq. (\ref{eq9full}) predicts that $\delta$ is a decreasing
function of the interaction strength $c_{FS}$. Now,
as emphasized for example by the Laplace estimate of the contact angle, 
$\cos\theta = -1 + 2\rho_S c_{FS} /\rho_F c_{FF}$, the contact angle may
be interpreted as a ``measure'' of the fluid-substrate interaction 
strength $c_{FS}$. In particular one expects the fluid to approach a 
non-wetting situation ($\cos \theta \rightarrow -1$), when $c_{FS}$ decreases
to zero. The previous equation, eq. (\ref{eq9full}), thus
predicts a strong increase of the slipping length $\delta$ when 
$\cos\theta \rightarrow -1$. 
In other words in the idealized situation of
a non wetting fluid, $\theta=\pi$, a perfect slip may be expected for the 
boundary condition of the fluid near the surface. The correct trend is observed
in our simulation results. This result is in agreement with several experimental
observations \cite{Churaev,Tolstoi}, reporting very large slipping lengths for nonwetting fluids.

\section{Conclusions}

Obviously the existence of such a large slippage effect 
should manifest itself in the dynamical properties of a liquid confined in a nanoporous medium. If one considers a single cylindrical capillary, a 
straightforward calculation in the Poiseuille geometry shows that 
the existence of slip on the boundaries increases the flow rate in the tube as compared
to the ``usual'' no-slip case by a factor $1+8\delta/h$ (with $h$ the pore diameter and
$\delta$ the slip length). Thus, in a porous medium, the effective
permeability $K_{eff}$, which relates according
to Darcy's law the flow rate to the pressure drop \cite{hydro}, is expected to increase by the same factor :
\begin{equation}
K_{eff}=K_0~\left( 1 + 8{\delta\over h}\right)
\label{permea}\
\end{equation}
where $K_0$ is the ''standard'' permeability, obtained within the no slip 
assumption ({\it i.e.}, when $\delta$ is zero). In a wetting situation,
$\delta$ is obtained to be very small and $K_{eff}\simeq K_0$. However, in
a nonwetting situation ($\theta \sim 140$ degrees), the slipping length 
$\delta$ may largely exceed the nanometric pore sizes $h$, so that
the effective permeability $K_{eff}$ is expected to be much larger than $K_0$ (say, more that one order of magnitude in view of the prefactors).

It can also be expected that the microscopic dynamics of the molecules 
could be rather different in a "nonwetting" medium, compared to what it is in 
the bulk or in a medium with
strong solid/liquid affinity. In fact, recent studies point towards
the importance of the surface treatment for the reorientation dynamics of 
small molecules in nanopores \cite{KREMER}. 
Correlating the wetting properties
with such microscopic studies seems to be a promising area for future research.

\thanks
This work was supported by the Pole
Scientifique de Mod\'elisation Num\'erique at ENS-Lyon, the CDCSP at the University of Lyon,
 the DGA and the French Ministry of Education  under contract
98/1776. We would like to thank E. Charlaix and P.-F. Gobin
for introducing us to this subject,
and Dr. S.J. Plimpton for making publicly available a parallel MD code  \cite{Plimpton},
a modified version of which was used in the present simulations. References
\cite{Churaev,Tolstoi} were pointed out to us by Dr. Remmelt Pit.

\begin{table}
\caption{Dependence of the surface tensions (in units of $\epsilon$) 
and contact angle $\theta$ (in degrees), on the tuning parameter $c_{FS}$. The liquid-vapour surface tension was determined 
independently to be $\gamma_{LV}=0.94~\epsilon$.} 
\vspace{6mm}
\begin{center}
\begin{tabular}{||c|c|c|c||} \hline
{\rule[-3mm]{0mm}{5mm}}
$c_{FS}$ & $\gamma_{SV}-\gamma_{LS}$ & $\cos(\theta)$ &  $\theta$   \\ \hline 
0.5& -0.71&-0.74&137    \\\hline
0.6& -0.65&-0.68&133    \\\hline
0.7& -0.50&-0.52&121    \\\hline
0.8& -0.35&-0.36&111    \\\hline
0.9& -0.16&-0.17&99     \\\hline
\end{tabular}
\end{center}
\end{table}

\begin{figure}
\psfig{file=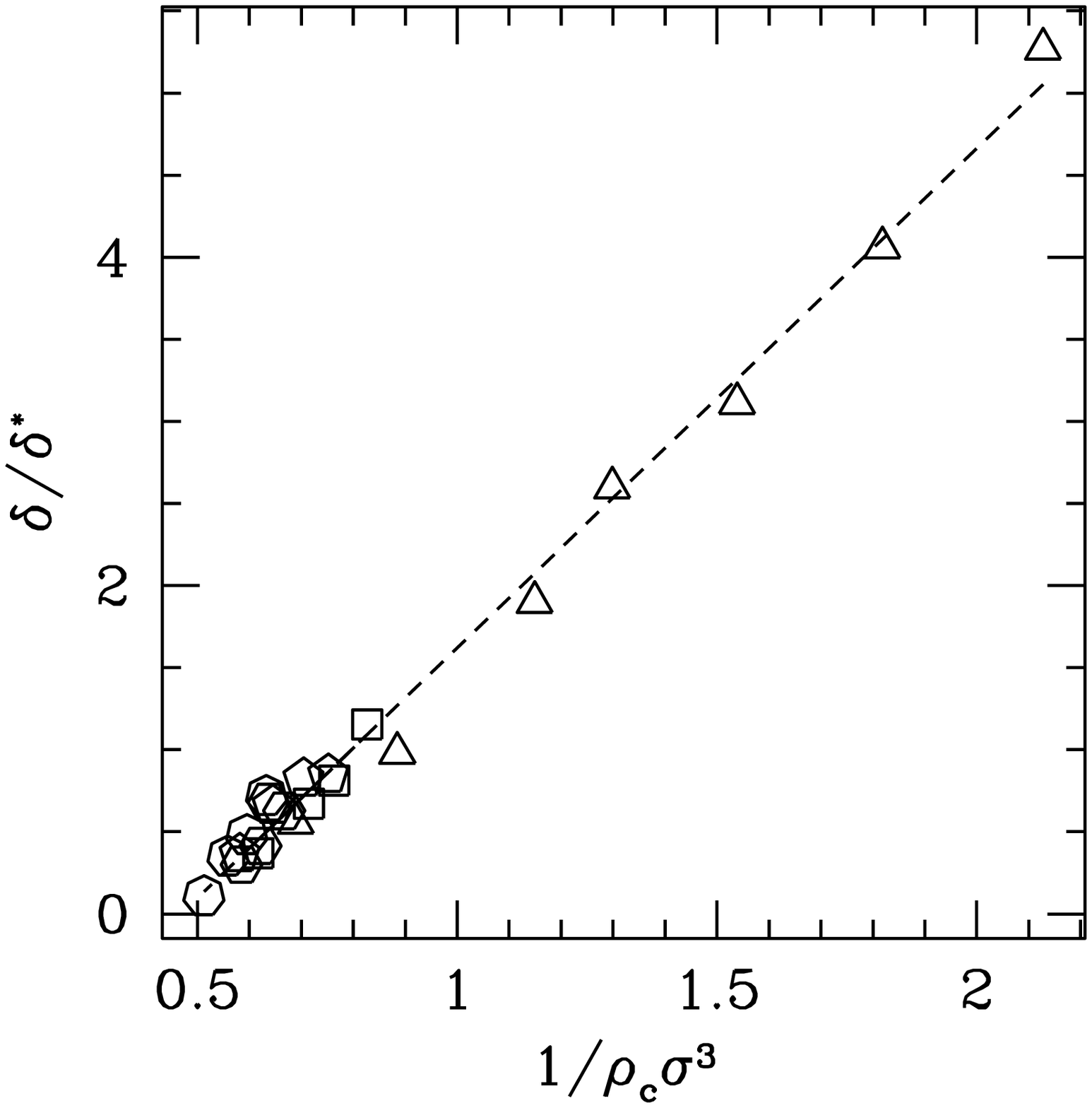,width=7.5cm,height=7.5cm}
\vspace{6mm}
\caption{Normalized slipping length 
$\delta/\delta^{*}$, with $\delta^{*}=\sigma D^{*}_{q_\parallel}/ 
{S_1(q_{\parallel}) c_{FS}^2}$,
as a function of the inverse contact density $1/\rho_c\sigma^3$. In this
plot, a linear dependence is expected according to the theoretical 
prediction, eq. (\ref{eq9}). The dashed line is a least-square fit of the
numerical datas, with slope $\alpha=3.04$ and shift in inverse density
$1/\rho_{shift}\sigma^3=0.47$.
}
\label{fig6}
\end{figure}


\begin{references}
\bibitem{note1} We mention that in some numerical simulations, purely repulsive interactions between the fluid and the substrate were considered. 
In that case, however, the pressure of the fluid is very high, and
the properties of the confined fluid is very similar to what is obtained for 
a wetting fluid (with attractive interactions to the substrate) at
a lower pressure.
\bibitem{note2} The contact angle of mercury on glass is typically 
140 degrees.
\end{references}
\end{document}